\documentclass[12pt]{article}

\usepackage{newtxtext,newtxmath}
\usepackage{graphicx}
\usepackage[letterpaper,margin=1in]{geometry}
\usepackage{scicite}
\usepackage{url}

\usepackage{authblk}
\usepackage{color}

\usepackage{ulem}
\usepackage[mathlines]{lineno}
\usepackage[english]{babel}
\usepackage[latin1]{inputenc}

\renewenvironment{abstract}
	{\quotation}
	{\endquotation}
\date{\today}

\makeatletter
\renewcommand{\fnum@figure}{\textbf{Figure \thefigure}}
\renewcommand{\fnum@table}{\textbf{Table \thetable}}
\makeatother

\def\scititle{Discovery of an intrinsic non-Hermitian phase transition in a bulk condensed-matter system}
\title{\bfseries \boldmath \scititle}

\author{

	Jingwen Li$^{1}$,
	Michael Turaev$^{2}$,
	Masakazu Matsubara$^{3,4,5}$,
    Kristin Kliemt$^{6}$,
    Cornelius Krellner$^{6}$,
    Shovon Pal$^{7}$,
    Manfred Fiebig$^{1,\ast}$,
    Johann Kroha$^{2,8,\ast}$\and
    
	\small$^{1}$Department of Materials, ETH Zurich, Vladimir-Prelog-Weg 4, 8093 Zurich, Switzerland.\and
	\small$^{2}$Physikalisches Institut and Bethe Center for Theoretical Physics, University of Bonn, 53115 Bonn, Germany.\and
	\small$^{3}$Division of Materials Physics, Graduate School of Engineering Science, The University of Osaka, Toyonaka, Osaka 560-8531, Japan.\and
	\small$^{4}$Department of Physics, Tohoku University, Sendai 980-8578, Japan.\and
    \small$^{5}$PRESTO, Japan Science and Technology Agency (JST), Kawaguchi 332-0012, Japan.\and
	\small$^{6}$Physikalisches Institut, Goethe-Universit\"{a}t Frankfurt, 60438 Frankfurt, Germany.\and
	\small$^{7}$School of Physical Sciences, National Institute of Science Education and Research, An OCC of HBNI, Jatni, 752 050 Odisha, India.\and
	\small$^{8}$School of Physics and Astronomy, University of St.\,Andrews, North Haugh, St.\,Andrews, KY16 9SS, United Kingdom.\and
    
	\small$^\ast$Corresponding author. Email: manfred.fiebig@mat.ethz.ch, jkroha@uni-bonn.de
}

\begin{document}

\maketitle

\begin{abstract} \bfseries \boldmath   %130 words -- ok at this stage (125 permitted)
Across regular phase transitions, systems remain in thermal equilibrium. However, when a system is driven far from equilibrium, non-Hermitian phase transitions may arise where the dynamical behavior rather than steady properties undergo a qualitative change at a critical, so-called exceptional point. We experimentally realize a non-Hermitian phase transition in a bulk condensed-matter system. Optical excitation creates charge carriers in ferromagnetic EuO. In a temperature-dependent interplay with the Hermitian transition to ferromagnetic order, a non-Hermitian change of the relaxation dynamics occurs, manifesting in our time-resolved reflection data as the transition from biexponential real to single-exponential complex decay. Our theory models this behavior and suggests that non-Hermitian phase transitions may generically emerge in bulk condensed matter.
\end{abstract}

Systems that undergo a phase transition typically remain in thermodynamic equilibrium at all times, leading to a lasting change in ground state properties. A well-known example is the emergence of spontaneous magnetization at the transition between the paramagnetic and ferromagnetic phases. To a good approximation, this holds even when new states are transiently stabilized on short timescales, for example, when pulsed lasers are used to control magnetically ordered states in the field of ultrafast magnetization dynamics \cite{Beaurepaire1996, Fausti2011, Nova2019, Sie2019, Rasing2010, Yang2023Terahertz}. Driving a system truly away from thermodynamic equilibrium could therefore create qualitatively new states of matter that are characterized by their dynamics rather than their ground-state properties and cannot be realized otherwise. This leads us to the subject of non-Hermitian phase transitions (NHPTs) \cite{Ashida2020}. In contrast to equilibrium systems, which are invariant with respect to the displacement in or, more specifically, reversal of time as a fundamental law of nature, excitation and dissipation break time-reversal symmetry. This usually manifests itself in an asymmetry between the drive and loss terms. The matrix connecting the response of a system's properties to the driving fields becomes non-Hermitian, its eigenvalues in general complex, and its eigenvectors non-orthogonal. Whereas in an equilibrium phase transition the free energies of two static phases (or, at zero temperature, the energies of two quantum states) become equal, a NHPT occurs when two real eigenvalues of a nonequilibrium response matrix $\chi$ become equal and change into a single complex-conjugate pair. Formally, this can be expressed by eigenvalues $\gamma_{1,2}$ of the form $\gamma_{1,2}=x\pm\sqrt{y}$, where $y$ changes its sign from ``$+$'' to ``$-$'' at the  exceptional point (EP) \cite{Heiss2012}. In consequence, the corresponding eigenvectors coalesce, which means physically that the two modes of the system dynamics collapse into a single one. This is accompanied by a qualitative change in dynamical behavior with enhanced sensitivity to the variation of external parameters near the EP. Therefore, a NHPT generally represents a transition in the dynamics of a system rather than in its equilibrium properties, where the latter would be described by a Hermitian Hamiltonian on the microscopic level \cite{Erdamar2024}. It is this dynamical behavior that has generated considerable interest in NHPTs. Most of its studies are theoretical \cite{Hanai2019,Diehl2024}, and experimental realizations have focused on single-spin or nanostructured magnetic \cite{Wu2019,Liu2019}, optical \cite{Guo2009,Rueter2010,Li2023,Peng2016}, ultracold-atomic-gas \cite{Li2019a,Li2019, Pan2019,Wang2024}, or photon-condensate systems \cite{Oeztuerk2021}, all of which are engineered. The possibility of an intrinsic NHPT in a homogeneous bulk solid-state system has not even been discussed because it was considered infeasible to establish the required extreme off-equilibrium  conditions in the bulk.

Here, we report on the observation of an intrinsic NHPT in the bulk ferromagnetic semiconductor europium monoxide (EuO) after optical excitation of electrons from the $4f$ valence to the $5d$ conduction band of the Eu ions. Time-resolved pump-probe measurements of the linear reflectivity reveal the characteristic change of the ensuing relaxation dynamics from two real to a single complex-conjugate decay rate. The EP of this transition is observed at $\approx 85$~K and therefore distinctly higher than the Curie temperature $T_{\rm C}$ of the Hermitian phase transition to ferromagnetic order at 69~K. Our theory models this behavior and demonstrates that, as a unique aspect of many-body systems, a Hermitian and a NHPT may be linked to each other as originating from the same microscopic coupling mechanism, although representing very different physical phenomena occurring at different critical temperatures. We finally propose, based on the common ingredients of our theoretical model, that NHPTs should be allowed for a large class of condensed-matter systems, where they can be exploited to sensitively control bulk dynamic properties. 
\subsection*{Signatures of a non-Hermitian phase transition in bulk condensed-matter}
EuO exhibits a rock-salt structure and has a semiconductor gap of $1.2$~eV. The ferromagnetic order of the Eu $4f^7$ ions is mediated by virtual magnetic polarons involving charge-carrier fluctuations in the spatially extended Eu $5d(t_{2g})$ orbitals that interconnect the tightly bound Eu $4f^7$ magnetic moments \cite{Mauger1977}. Doping charge carriers into the $5d(t_{2g})$ conduction orbitals by chemically substituting Eu with Gd atoms further enhances $T_{\rm C}$ \cite{Mairoser2010,Arnold2008} through the Ruderman-Kittel-Kasuya-Yosida interaction \cite{Coey2010} mediated by the conduction electrons. Alternatively, $4f^7 \to 5d(t_{2g})$ photodoping leads to a transient strengthening of the magnetic order \cite{Masakazu2015}. Both effects underpin the importance of the magnetic coupling $J_{df}$ between the $4f$ local moments and the $5d$ conduction spins in EuO. The ferromagnetic transition is accompanied by a pronounced Stoner band splitting, which shifts the spectral weight below the Fermi energy and, thus, makes the material metallic with nearly completely spin-polarized electrons at low temperature \cite{Steeneken2002,Arnold2008}. 

We use EuO single crystals grown by the gradient cooling technique with a Bridgman setup \cite{Reed1971}. The crystals are cut to yield a [100]-oriented surface, lapped with SiC films to a thickness of 280~$\mu$m, and polished with Al$_2$O$_3$ films in ethanol. In a pump-probe experiment sketched in Fig.~\ref{fig:time_trace}A and further detailed in \cite{SM}, a 120-fs pump pulse of 1.55~eV, generated by an amplified Ti:sapphire laser system, promotes electrons from the Eu $4f$ valence to the $5d(t_{2g})$ conduction band. Note that the total spin $S=7/2$ is conserved in this transition (Fig.~\ref{fig:time_trace}E). We then record $\Delta R(t)/R$, the change in reflectivity with respect to the non-pumped system after a delay $t$. For this we use 120-fs laser pulses at 1.31~eV, which are emitted by a downstream optical parametric amplifier and incident under an angle of 45$^{\circ}$. These pulses probe the dipole-allowed, spin-conserving electronic Eu $4f \to 5d(t_{2g})$ transition at 1.29~eV resonantly (Fig.~\ref{fig:time_trace}E). Figure~\ref{fig:time_trace}B shows the corresponding reflectivity time traces after optical pumping between 10~K and 296~K.

Across this temperature range, the relaxation behavior exhibits a drastic qualitative change. At all temperatures, we observe an initial increase in the reflectivity that peaks within the duration of the laser pulse. At the lowest temperatures, subsequent relaxation exhibits a rapid and continuous decay of the reflectivity change on a time scale of about 1~ps, followed by a much slower second relaxation (Fig.~\ref{fig:time_trace}C). In the ferromagnetic phase, the normalized reflectivity change $\Delta R/R$ settles to a positive value, indicating that the optically induced conduction-band population is energetically stabilized by the aforementioned Stoner shift of the majority conduction band below the Fermi energy \cite{Steeneken2002,Arnold2008} at least throughout our observed time range of 20~ps. Toward the highest temperatures, however, the initial reflectivity peak is followed by a single rapid decay until $\Delta R$ becomes negative before approaching the original value at $\Delta R=0$ (Fig.~\ref{fig:time_trace}D). For the pump fluence of 100~$\mu$J/cm$^{2}$ used in the experiment, the transition from the all-positive to the partly negative evolution of $\Delta R/R$ occurs in the paramagnetic phase, at a threshold temperature $T^* = (84.4\pm 2.6)$~K or $(87.8\pm 1.6)$~K in two independent fits that are discussed in \cite{SM}. The values are consistent and lie clearly above $T_{\rm C}$. In the following, we use $T^* \approx 85$~K, as the threshold temperature.

We point out that common phenomena known to possibly cause a negative dynamic response cannot explain the behavior in Fig.~\ref{fig:time_trace}D. Among the multitude of effects that might come to mind are oscillations induced by coupling to coherent lattice vibrations \cite{Sun2017}, Auger recombination \cite{But2019, Sharma2024}, band-filling effects in semiconductors \cite{Sabbah2002, Prabhu2004, Cicco2011, Cicco2020} temperature or strain-dependent band shifts \cite{Jnawali2020}, three-temperature-model relaxation \cite{Shim2020}, lattice heating, excitation of optical phonons, or paramagnetic excitations \cite{Miyamoto2018}. Those and other effects are discussed and ruled out in \cite{SM}. In addition, none of these effects can explain (i)  the very specific temperature and pump-fluence dependencies of the relaxation rates shown below and (ii) the presence of a critical temperature $T^*$ for the onset of these dependencies.

\subsection*{Understanding the non-Hermitian phase transition in EuO}

\begin{figure}
    \centering
   \includegraphics[width=\textwidth]{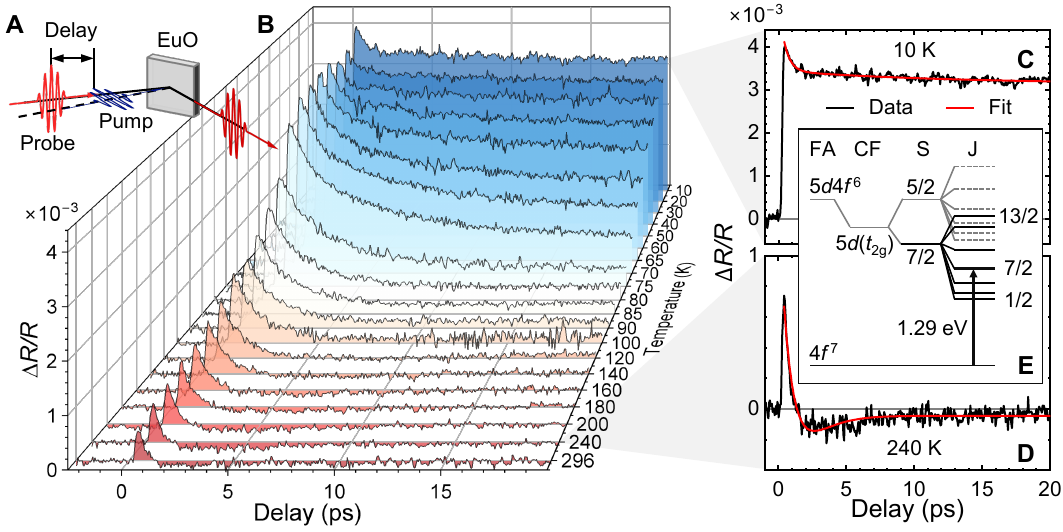}
    \caption{\textbf{Photoinduced reflectivity change in EuO.}
    {\bf (A)} A 120-fs laser pulse (blue) with an energy density of 100~$\mu$J/cm$^2$ and a photon energy of 1.55~eV drives the system out of equilibrium and excites electrons from the Eu~$4f$ shell to the Eu~$5d(t_{2g})$ orbitals. Time-delayed probe pulses (red) of $1.31$~eV induce resonant recombination of $5d(t_{2g})$ electrons with holes in the Eu~$4f$ shell; see (E). The photons emitted by this recombination contribute to the reflectivity change. {\bf (B)} Relative change of reflectivity with respect to the non-pumped system as a function of delay for temperatures from 10 to 296~K. The NHPT occurs when the relaxation dynamics changes from {\bf (C)} biexponential positive to {\bf (D)} sign-changing; see text. {\bf (E)} Electronic level scheme around the semiconductor gap of EuO \cite{Wang1986} for the free atom (FA), the crystal field (CF), the spin ($S$), and the total angular momentum ($J=L+S$)}
    \label{fig:time_trace}
\end{figure}

\begin{figure}
    \centering
   \includegraphics[width=\textwidth]{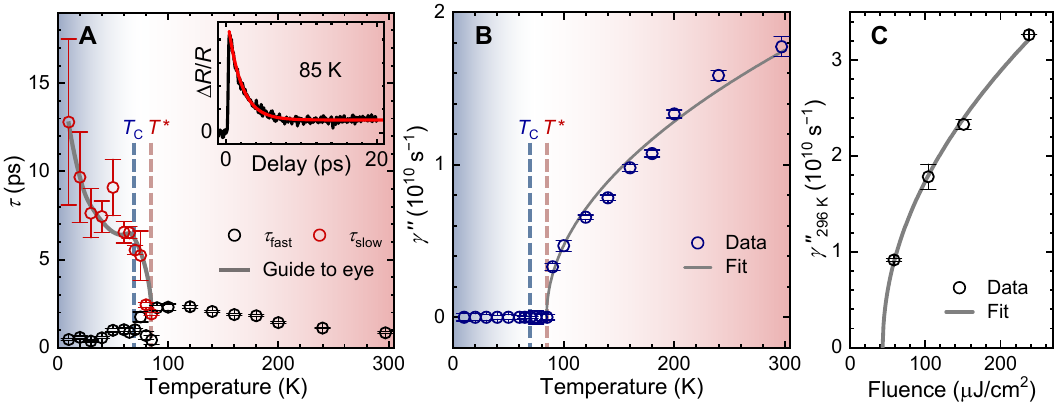}
    \caption{\textbf{Hallmarks of a non-Hermitian phase transition.} {\bf (A)} The biexponential relaxation times $\tau_{\rm fast}$, $\tau_{\rm slow}$ and {\bf (B)} the imaginary part $\gamma''$ of the relaxation rate $\gamma$ are shown as a function of temperature, extracted from the time traces of Fig.~\ref{fig:time_trace}B. The appearance of a non-zero value $\gamma''$ at the critical temperature of $T^*\approx 85$~K marks the exceptional point (EP). The inset shows that at the EP, the fit of a single exponential decay (red line) describes the relaxation behavior. Note the cusp-like feature in $\tau_{\rm fast}$ and $\tau_{\rm slow}$ at the ferromagnetic ordering temperature $T_{\rm C}=69$~K. {\bf (C)} Relaxation parameter $\gamma''$ as a function of pump fluence $P$. The fitted scaling behaviors $\gamma'' \sim \sqrt{T-T^*}$ above the EP for fixed pump fluence and $\gamma''\sim \sqrt{P-P^*}$ at fixed temperature are in excellent agreement with theory and pose unambiguous signatures of a NHPT (see text and Fig.~\ref{fig:theo_rates}). At $T=296$~K, the threshold pump fluence for the onset of complex relaxation is $P^* \approx 42~\mu$J/cm$^2$.}
    \label{fig:exp_rates}
\end{figure}

Having thus excluded known physical origins for the measured pump-induced transient reflectivity decrease, we now show that our observed response exhibits, in fact, the unique signature of a NHPT. For further analysis, we extract the relaxation rates $\gamma_1$, $\gamma_2$ at $T<T^*$ by a biexponential fit to the experimental data according to
\begin{equation}
  \frac{\Delta R(t)}{R} = C_1 {\rm e}^{-\gamma_1 t} + C_2 {\rm e}^{-\gamma_2 t} + \Delta_0,
\end{equation}
with real $\gamma_{1,2}$, $C_{1,2}$ and $C_{1,2}>0$. $\Delta_0$ is the long-term-stable Stoner-shift-induced offset, which approaches zero toward $T^*$. To describe the negative response at $T>T^*$, there are two options. First, we can assume that either $C_1$ or $C_2$ becomes negative. As we show in \cite{SM}, the associated fits do not converge well, especially in the critical temperature range, where the negative contributions emerge, irrespective of the chosen pump fluence. In addition, as discussed above, we have excluded contributions from processes that involve negative exponential behavior for physical reasons. Second, we allow for complex relaxation rates and set $\gamma_{1,2}=\gamma'\pm {\rm i} \gamma''$ and $C_{1,2}=C_0 {\rm e}^{\pm {\rm i}\varphi_0}$, with real values $\gamma'$, $\gamma''$, $C_0$, and $\varphi_0$, so that real values for $\Delta R(t)/R$ are retained. This fit converges well at all temperatures above $T^*$ and for different laser fluences, and the results for the relaxation time(s) $\tau={\rm Re}(\gamma)^{-1}$ and the relaxation parameter $\gamma'' = {\rm Im}(\gamma)$ are shown in Fig.~\ref{fig:exp_rates}. As a detailed fit error analysis (Fig.~\ref{fig:biexponential+complex}) in \cite{SM} shows, we observe biexponential relaxation with two relaxation times for $T<T^*$, whereas for $T>T^*$ the dynamical behavior collapses to a single relaxation mode with regions of positive and negative measured data caused by the complex nature of the relaxation rate. Degeneracy occurs at $T^*$ where a single exponential (with an additional powerlaw factor) describes the relaxation according to the inset of Fig.~\ref{fig:exp_rates}A. This collapse at a specific temperature $T^*$ is most striking in its resemblance to a system undergoing a NHPT. In particular, whereas several physical processes could in principle account for a negative signal like in Fig.~\ref{fig:time_trace}D and have been evaluated (and ruled out) as stated above, none of these processes except for the NHPT can explain the very distinctive square-root dependence of the relaxation parameter on temperature and pump fluence $P$ shown in Figs.~\ref{fig:exp_rates}, B and C, respectively. 

In the following, we present a theoretical model that explains the occurrence of a NHPT in EuO in a natural way and describes all the temperature- and fluence-dependent features on the same footing, including the cusp visible in Fig.~\ref{fig:exp_rates}A at the Curie temperature $T_{\rm C}$. To do this, we first identify the dynamical variables active in our material. The change in reflectivity after optical excitation is proportional to the density of photoexcited charge carriers in the conduction band, which recombine with holes in the valence band by the emission of a photon. Excitations of this type, Wannier excitons, have long been studied in EuO \cite{Mitani1975}, and an associated absorption spectrum is shown in Fig.~\ref{fig:OD_spectrum}. The biexponential relaxation after optical pumping at low temperature is in good agreement with the fact that there are two types of electronic excitations. These are ``bright'' (optically excitable) and ``dark'' (non-radiatively decaying) excitons, both formed as a bound state of a $5d(t_{2g})$ electron and a $4f$ hole in the valence band \cite{Loh2017}. Furthermore, it is known that the conduction-electron spin is strongly coupled to the Eu $4f^7$ local moments via a Heisenberg exchange coupling $J_{df}$ (see the introduction on EuO above). 

\begin{figure}
    \centering
   \includegraphics[width=\textwidth]{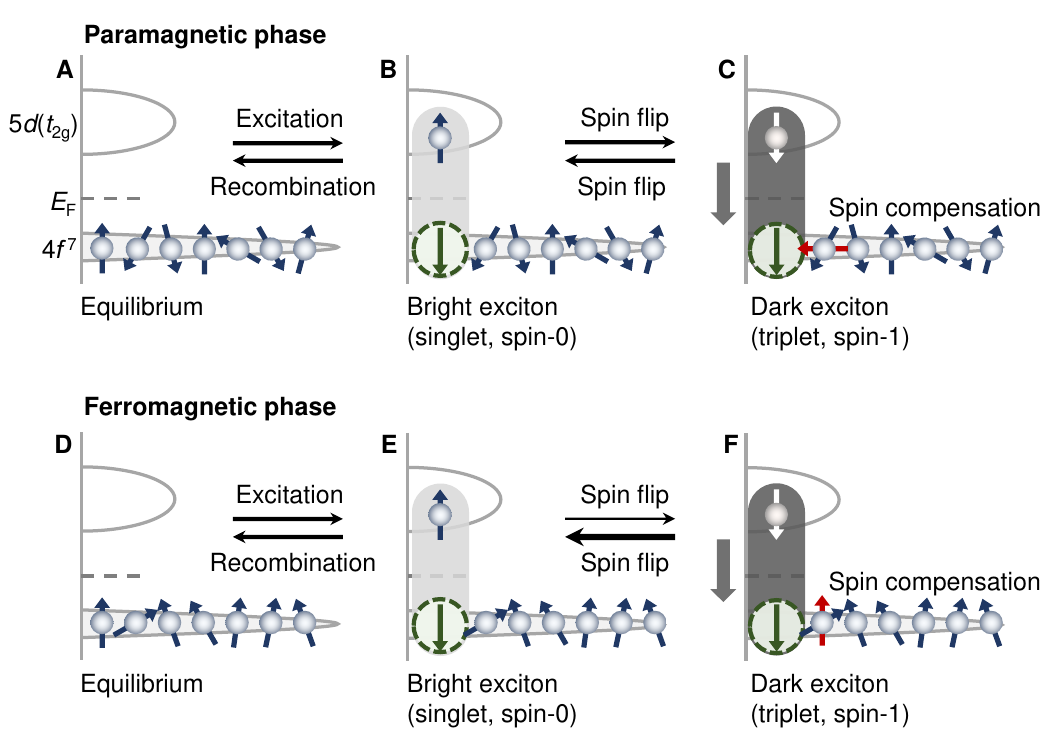}
    \caption{\textbf{Schematic of the coupled spin-charge dynamics in EuO.} {\bf (A} to {\bf C)}. Paramagnetic phase. {\bf (A)} The electronic structure of EuO is comprised of a Eu~$4f^7$ valence band and a $5d(t_{2g})$ conduction band, with the Fermi level $E_{\rm F}$ in the gap of $1.2$~eV. {\bf (B)} Photoexcitation of electrons at 1.55~eV into the Eu $5d(t_{2g})$ orbitals generates bright excitons which can recombine by the emission of a photon at $1.31$~eV. This is recorded in the experiments as the positive (normalized) reflectivity change $\Delta R/R>0$. {\bf (C)} Before recombination, a bright exciton (spin-0) can be transformed into a dark exciton (spin-1) by flipping the electron spin by interaction with a Eu~$4f$ (7/2) magnetic moment. The total spin is conserved by the $4f$ magnetic moments changing their orientation accordingly (blue-to-red orientation). The spin-1 excitons cannot directly recombine which reduces the density of photoemitting charge carriers and thus leads to the negative contribution to $\Delta R/R$. {\bf (D} to {\bf F)}, Ferromagnetic phase. The spontaneous orientation of the $4f$ spins induces an asymmetry between bright-to-dark and dark-to-bright exciton transitions (spin-flip arrows of different thickness). Its onset leads to a cusp in the relaxation times at the Curie temperature that is visible in Figs.~\ref{fig:exp_rates}A and \ref{fig:theo_rates}A.}
    \label{fig:schematic}
\end{figure}

Putting this together, an ultrashort optical laser pulse excites electrons across the semiconductor gap of 1.2~eV without spin flip (Fig.~\ref{fig:time_trace}E), thus creating bright excitons which can recombine by the emission of a photon and induce the measured reflectivity change (Fig.~\ref{fig:schematic}, A and B, D and E). Alternatively, a conduction electron can flip its spin with a local Eu $4f$ moment through the coupling $J_{df}$, thus transforming a bright spin-0 exciton into a dark spin-1 exciton and vice versa (Fig.~\ref{fig:schematic}, C and F). The spin-1 exciton cannot recombine by emitting a photon via the spin-conserving dipole decay, thus reducing the reflectivity below its equilibrium value ($\Delta R/R<0$). During the bright-dark exciton transformation, the orientation of the randomly oriented Eu $4f$ moments is also altered. Because of the different energies of spin-0 and spin-1 excitons (see Fig.~\ref{fig:time_trace}E), the bright-dark transformation processes are dissipative and thus energy-non-conserving. The energy difference is absorbed or provided by a thermal bath of phonons at cryostat temperature $T$. This induces an asymmetry between the bright-dark and dark-bright transformation amplitudes, $\Gamma^+$, $\Gamma^-$, which breaks the time-reversal symmetry and makes these amplitudes temperature dependent, as formally derived by the Lindblad formalism \cite{Petruccione2010}.

\begin{figure}
    \centering
   \includegraphics[width=0.67\textwidth]{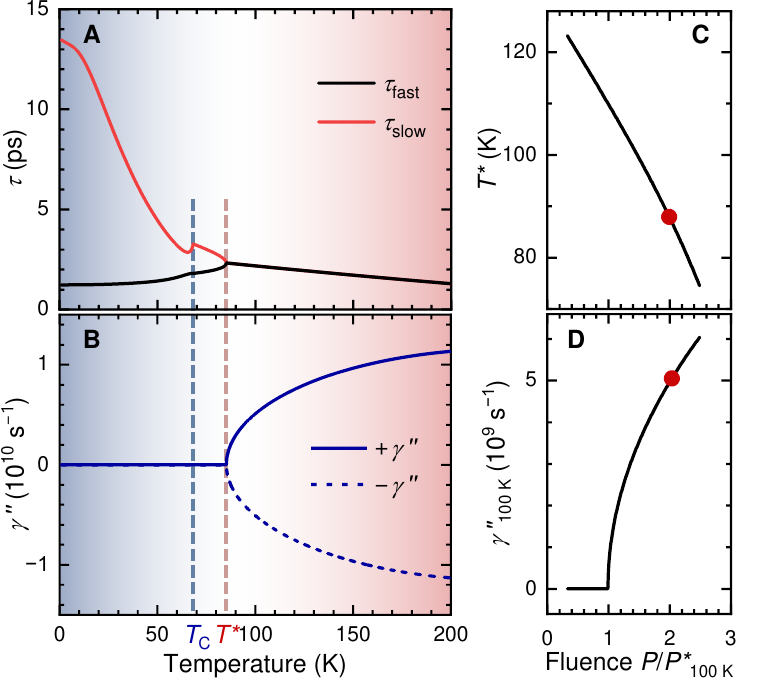}
    \caption{\textbf{Theoretical results.}
    {\bf (A} and {\bf B)} Relaxation times $\tau_{\rm fast}$, $\tau_{\rm slow}$ as well as the imaginary part $\gamma''$ of the relaxation rate $\gamma$ are shown as a function of temperature, exhibiting an exceptional point at $T^*$. Parameters used for the calculations: see Table S1 in \cite{SM}. The qualitative agreement with the experimental observations (Fig.~\ref{fig:exp_rates}), including the cusp at $T_{\rm C}$, strongly supports our model and the presence of a NHPT. {\bf (C)} Dependence of the EP at $T^*$ on the pump fluence $P$ as calculated from the Lindblad dynamical equations, see \cite{SM}. {\bf (D)} The dependence of $\gamma''$ on $P^*$, here calculated at 100~K, scales as $\sqrt{P-P^*}$ beyond the critical pump fluence $P^*$, again in agreement with the experimental observation in Fig.~\ref{fig:exp_rates}. The red dots in (C) and (D) indicate the pump fluence used for the calculations in (A) and (B).} 
    \label{fig:theo_rates}
\end{figure}

In the paramagnetic phase, this leads to a set of two coupled rate equations for the densities of bright and dark excitons, $n_{\rm b,d}(t)$. The equations are parameterized by a real-valued, asymmetric $2\times 2$ matrix $\chi (T)$, see Eq.~\ref{eq:dynamical_eq1} in \cite{SM}. Its eigenvalues are the relaxation rates $\gamma_1$ and $\gamma_2$ of the bright-dark--exciton-coupled system, which are the solutions of a quadratic equation of the form
\begin{equation}
  \gamma_{1,2}= \left[\gamma_{\rm b} + \gamma_{\rm d} \pm \sqrt{(\gamma_{\rm b}-\gamma_{\rm d})^2 -4 \omega_0^2}~\right]/2. 
  \label{eq:eigenvalue}
\end{equation}
Here, $\gamma_{\rm b}$, $\gamma_{\rm d}$ are the diagonal elements of $\chi$, that is, the bright- and dark-exciton decay rates without mutual coupling, and $\omega_0^2$ is the modulus of the product of the off-diagonal elements of $\chi$. If there was no dissipation, $\omega_0$ would be the natural frequency of the bright-dark--exciton-coupled system. Figures~\ref{fig:theo_rates}, A and B, show the relaxation times $\tau_{\rm fast}$, $\tau_{\rm slow}$ and the relaxation parameter $\gamma''$ as the inverse real parts and as the imaginary part of these eigenvalues, respectively, whose temperature dependence is inherited from the transition amplitudes $\Gamma^{\pm}(T)$. The EP is reached when the argument of the square root in Eq.~\ref{eq:eigenvalue}, crosses zero at the critical temperature $T^*$ of the NHPT. As the possibly most striking and unique feature associated with the NHPT, Eq.~\ref{eq:eigenvalue} predicts that the imaginary part of the relaxation rate behaves as $\gamma'' \sim\sqrt{T-T^*}$ for $T\geq T^*$, see Fig.~\ref{fig:theo_rates}B. This is because the argument of the square root crosses zero linearly at the critical values. The fact that the experiment shows precisely this behavior (see Fig.~\ref{fig:exp_rates}B), which is not expected from other physical phenomena, is an unambiguous confirmation that the optically pumped EuO exhibits a NHPT with the sample temperature as the control parameter.  

In the ferromagnetic phase, the $4f$ spin orientation acquires, in addition to the time-dependent variation, a non-zero, static equilibrium value, the magnetization, which has the temperature dependence $M(T) = M_0\sqrt{1-T/T_{\rm C}}$. This temperature dependence manifests itself in the relaxation times as a cusp visible in Fig.~\ref{fig:theo_rates}A near the Curie temperature $T_{\rm C}$. A similar feature appears in the experimental data, Fig.~\ref{fig:exp_rates}A, at $T_{\rm C}=69$~K, which lends additional support to the correctness of our model and, hence, for the existence of the NHPT. The ferromagnetic transition and the NHPT are closely related because they are caused by the same microscopic interaction, namely the spin coupling $J_{df}$ between Eu $4f$ moments and electrons in the $5d(t_{2g})$ orbitals. However, they represent fundamentally different physical phenomena. These are the phase transitions in the ground-state properties (ferromagnetic $\leftrightarrow$ paramagnetic) and in the relaxation dynamics (biexponential $\leftrightarrow$ complex-conjugate) that generally occur at different critical temperatures, $T_{\rm C}$ and $T^*$, respectively. 

We further note that the relaxation parameter $\gamma''$ is not a fixed intrinsic material parameter but is of dynamic origin, determined by the interplay of bright-dark exciton transformation and dissipation. The pump fluence $P$ controls how far the system is driven into nonequilibrium and determines the initial photoexcited exciton densities $\overline{n_{\rm b}}$, $\overline{n_{\rm d}}$. Our model thus predicts that with increasing pump fluence $T^*$ shifts towards lower values (Fig.~\ref{fig:theo_rates}C). At a given temperature, the imaginary part of the decay rate, therefore, scales as $\gamma''\sim \sqrt{P-P^*}$ across the NHPT at the critical pump fluence $P^*$ (Fig.~\ref{fig:theo_rates}D). Again, this prediction is supported by our experiments; see Fig.~\ref{fig:exp_rates}C. We then use Figs.~\ref{fig:theo_rates}C and D to construct the schematic 3D phase diagram (Fig.~\ref{fig:phase_diagram}) of the dynamical space of the NHPT in EuO.

\subsection*{Discussion and outlook}

\begin{figure}
    \centering
   \includegraphics[width=0.8\textwidth]{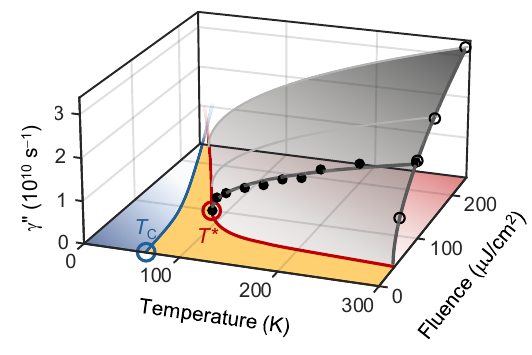}
    \caption{\textbf{Schematic phase diagram of the coexistence of NHPT and ferromagnetic Hermitian order in EuO.} The full or open data points represent the measured relaxation parameter $\gamma ''$ with fits (grey lines) as shown in Figs.~\ref{fig:exp_rates}, B and C, respectively. The blue line shows the qualitative dependence of the Curie temperature $T_{\rm C}$ on the pump fluence. Upon increasing this fluence from zero, the conduction-electron-mediated ferromagnetic coupling between the Eu~$4f$ moments transiently increases on the picosecond time scale without, however, significantly changing $T_{\rm C}$ \cite{Masakazu2015}. Only for stronger pump fluences is $T_{\rm C}$ reduced owing to the increasing inelastic processes. The red line shows the dependence of $T^*$ on the pump fluence. It decreases with increasing fluence as derived from our theory (see Fig.~\ref{fig:theo_rates}C) such that it may eventually coincide with $T_{\rm C}$, creating a critical EP with altered critical fluctuations and enhanced sensitivity to external parameters. In the limit of small fluence, $T^*$ eventually levels off so that a minimum fluence is required to reach the EP. Colored areas: yellow -- biexponential relaxation, red -- complex-conjugate relaxation, blue -- ferromagnetic phase, grey -- behavior of $\gamma ''$ in the temperature/laser-fluence plane. Colored circles: red -- temperature $T^*$ for the laser pump fluence of $100~\mu $J/cm$^2$ used in Figs.~\ref{fig:time_trace} and \ref{fig:exp_rates}, blue -- equilibrium value of $T_{\rm C}$.}
    \label{fig:phase_diagram}
\end{figure}

To conclude, we have discovered a new dynamical state of matter in the exciton gas of a bulk solid-state material, europium oxide, which manifests itself in complex relaxation behavior detected by our optical pump-probe experiments. The nonequilibrium phase diagram of EuO encompasses both a Hermitian phase transition to ferromagnetic order and, as reported here, a NHPT towards the dynamical phase.
Both transitions are mediated by the magnetic coupling of Eu $4f^7$ moments and $5d (t_{2g})$ electrons, but represent separate phenomena --- a phase transition between ground-state properties and a phase transition in the relaxation behavior --- which occur at different critical temperatures, $T_{\rm C}$ and $T^*$, respectively. We suggest that, based on Fig.~\ref{fig:phase_diagram}, both critical temperatures in EuO may be tuned via the pump fluence to coincide with characteristically altered critical fluctuations and enhanced sensitivity to parameter changes as a consequence \cite{Diehl2024}. This may be verified once we have found a way to realize the necessary fluence. Most importantly, none of the requirements of our theoretical model for the intrinsic bulk NHPT are specific to EuO. The necessary ingredients --- two coupled excited states or modes subject to dissipative dynamics --- are widely realized in condensed-matter systems. The general nature of our model therefore suggests that similar non-Hermitian phase transitions could occur across a wide range of materials.

\pagebreak

\pagebreak

\pagebreak

\pagebreak

\pagebreak

\section*{Acknowledgements} 
    We thank Nazia Kaya and Isabel Reiser for their support in the growth and characterization of the EuO crystals.  
\paragraph*{Funding:}
    This work was financially supported by the Swiss National Science Foundation through SNSF grants No. 200021\_178825, 200021\_219807, 200021\_215423 (J. L, M. F.), by the Deutsche Forschungsgemeinschaft (DFG) through the CRC/TR 185 (277625399) and the Cluster of Excellence ML4Q (EXC 2004/1-390534769) (M. T., J. K.), and through the CRC/TRR 288 (422213477, project A03) (C.K., K. K.), and by JSPS KAKENHI (Grants Nos. 24H00413 and 24H01639) and JST PRESTO (Grant No. JPMJPR23H9) (M. M.), and by DAE through the projects RIN4001 and RNI4011, and the startup support from DAE through NISER and SERB through SERB SRG through Project No. SRG/2022/000290 (S. P.).
\paragraph*{Author contributions:}
    J. L. performed the experiments after the initial reflectivity studies performed by M. M. J. L. and M. T. analyzed the data.  K. K. and C. K. produced and characterized the samples. S. P. and M. F. conceived and supervised the experiments. M. T. and J. K. developed the theory. All authors contributed to the discussion and interpretation of the experiment and to the completion of the manuscript.
\paragraph*{Competing interests:}
    The authors declare that they have no competing financial interests.
\paragraph*{Data and materials availability:}
    The data that support the findings of this article are openly available \cite{Zenodo-data}. Details for the material synthesis, data analysis, and theoretical modeling are in the supplementary materials \cite{SM}.

\subsection*{Supplementary materials}
Materials and Methods\\
Supplementary Text\\
Figures S1 to S6\\
Tables S1\\
References \textit{(46-\arabic{enumiv})}
%Movie S1\\
%Data S1

\newpage

\renewcommand{\thefigure}{S\arabic{figure}}
\renewcommand{\thetable}{S\arabic{table}}
\renewcommand{\theequation}{S\arabic{equation}}
\renewcommand{\thepage}{S\arabic{page}}
\setcounter{figure}{0}
\setcounter{table}{0}
\setcounter{equation}{0}
\setcounter{page}{1}

\begin{center}
\section*{Supplementary Materials for\\ \scititle}
	Jingwen Li,
	Michael Tuarev,
	Masakazu Matsubara,
    Kristin Kliemt,
    Cornelius Krellner,
    Shovon Pal,
    Manfred Fiebig$^{\ast}$,
    Johann Kroha$^{\ast}$\\
\small$^\ast$Corresponding author. Email: manfred.fiebig@mat.ethz.ch, jkroha@uni-bonn.de\\
\end{center}

\subsubsection*{This PDF file includes:}
Materials and Methods\\
Supplementary Text\\
Figures S1 to S6\\
Tables S1

\newpage

\subsection*{Materials and Methods}

\subsubsection*{Sample growth:} 
EuO single crystals were grown in a tungsten crucible enclosed in tantalum by combining high-purity Eu$_2$O$_3$ (99.9\%) and Eu (99.99\%) in a ratio of 1:3 as starting constituents. The Eu excess was necessary to suppress the formation of Eu$_3$O$_4$. Crystal growth was achieved by cooling the melt from 2280 to 1180~$^\circ$C over five hours in a high-frequency ``Arthur D. Little'' furnace using a Bridgman setup. See Ref. \cite{Reed1971} for details of the growth procedure. The specimen obtained had a volume of approximately 0.5 cm$^3$. Excess Eu was found on top of the grown crystal and immediately formed Eu oxide (Eu$_2$O$_3$) upon air contact. This Eu$_2$O$_3$ layer was removed by polishing, and subsequent Laue diffractometry showed that the grown sample was a large EuO single crystal with minor misorientation caused by small-angle grain boundary formation. Magnetization and specific heat measurements confirmed that the samples exhibit the corresponding physical properties known from the literature \cite{Ahn2005}. Prior to the optical measurements, all samples were hand-polished to obtain a fresh [100]-cut surface. 

\subsubsection*{Sample quality and characterization:}

\begin{figure}[ht]
    \centering
    \includegraphics[width=\textwidth]{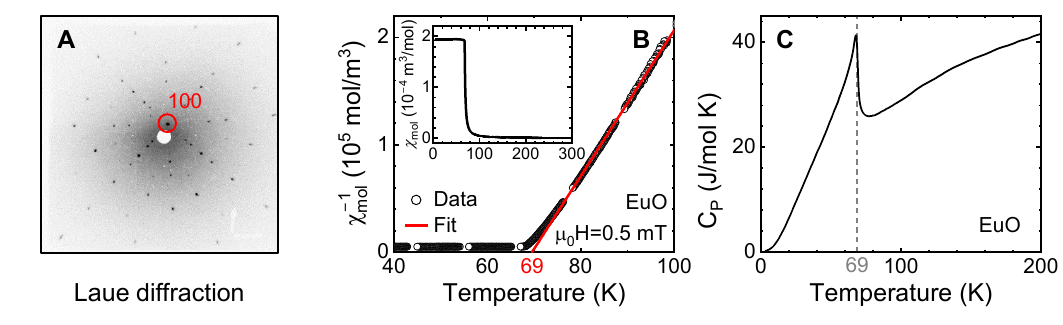}
    \caption{Characterization of our EuO bulk single crystals. \textbf{a}, The Laue pattern of the crystallographic $[100]$ direction shows the four-fold symmetry of the cubic lattice. \textbf{\rm b}, Inverse magnetic susceptibility of EuO measured in a 0.5\,mT magnetic field. The extrapolation of the linear fit leads to $T_{\rm C} \approx 69$~K. The inset shows the magnetic susceptibility from 1.8~K to room temperature. \textbf{c}, Zero-field specific heat.}
    \label{fig:material_characterization}
\end{figure}

We determine the composition of the sample by energy-dispersive X-ray spectroscopy, the crystal structure by powder X-ray diffraction, and the orientation of the as-grown EuO single crystals using a Laue diffractometer, M\"{u}ller Micro 91, with a tungsten anode. The typical Laue diffraction pattern (Fig.~\ref{fig:material_characterization}A) shows sharp reflexes that indicate high crystal quality of the samples. Furthermore, simulation of the pattern yields [100] cleavage planes. By indexing the peaks using the software ``OrientExpress", we confirm the rock-salt crystal structure and the lattice parameter ($a=5.1378$~\AA), which is consistent with the value in the literature \cite{Ahn2005, Belayev1974, Dillon1964}. 

We further checked the magnetic properties of our EuO samples by examining the magnetic susceptibility between 1.8\,K and 300\,K at $\mu_0H=0.01$\,T using a Quantum Design Physical Property Measurement System (PPMS). The sample was cooled in zero field prior to each measurement. The inset of Fig.~\ref{fig:material_characterization}B shows the magnetic susceptibility $\chi_{\rm mol}$ throughout the measured temperature range. The susceptibility $\chi_{\rm mol}$ exhibits an abrupt change at $T_{\rm C}=69~$ K, the value reported for the transition to the ferromagnetic state. The linear temperature dependence of $\chi^{-1}_{\rm mol}$ in the paramagnetic phase indicates the Curie-Weiss behavior. We determine the transition temperature by extrapolating the linear fit as $T_{\rm C} \approx 69$~K, which is precisely in agreement with the value in the literature \cite{McGuire1964,Mairoser2010}. This implies an undetectable low charge-defect density in our EuO samples, as $T_{\rm C}$ would sensitively shift to higher values even with small amounts of carrier doping $<0.1$\% \cite{Mairoser2010,Arnold2008}. This confirms that the conduction band is empty before the pump excitation. The near-constant value of $\chi^{-1}_{\rm mol}$ below $T_{\rm C}$ indicates a negligible coercivity in the susceptibility measurements, again in agreement with the literature \cite{Ahn2005}. The absence of a low-temperature increase of $\chi_{\rm mol}$ below 10~K \cite{Ahn2005} confirms the absence of compositions like Eu$_3$O$_4$, which would order antiferromagnetically at around 5~K. 

The ferromagnetic phase transition is also confirmed by the heat-capacity measurement between 1.8~K and 200~K in zero field using the PPMS, as displayed in Fig.~\ref{fig:material_characterization}C. The pronounced $\lambda$-like peak centered at 69~K matches the second-order transition from the ferromagnetic to the paramagnetic phase.

\begin{figure}[ht]
    \centering
    \includegraphics[width=0.5\textwidth]{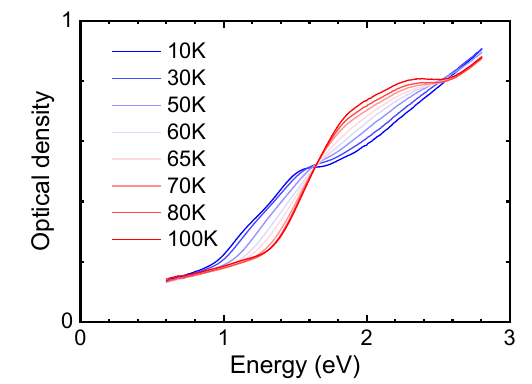}
    \caption{Optical-density spectrum of a bulk-like EuO film deposited on an YAlO$_3$ substrate and capped with amorphous silicon. The set of scans between 10 and 100~K was measured with unpolarized light. Adapted from Ref. \cite{Matsubara2010}.}
    \label{fig:OD_spectrum}
\end{figure}

The linear optical absorption of EuO is represented by the optical-density (OD) spectrum in Fig.~\ref{fig:OD_spectrum}, which was adopted from the literature \cite{Matsubara2010}. The spectrum reveals a pronounced temperature dependence of the OD. The crossing point at 1.7~eV and the opposite temperature dependence on either side reveal a Stoner splitting of the energy levels of the $5d(t_{2g})$ band (see the Fig.~\ref{fig:time_trace}E and the main text).

\subsubsection*{Pump-probe experiments:}

The samples were mounted in a Janis SVT-400 helium reservoir cryostat. A \textit{p}-polarized pump pulse at a photon energy of 1.55-eV, a pulse length of 120-fs, and a fluence of 100~$\mu$J/cm$^2$ was used to drive the samples out of equilibrium. A time-delayed 1.31-eV, 120-fs probe pulse with a fluence of around 30~$\mu$J/cm$^2$ was used to examine the pump-induced reflectivity change $\Delta{R}$ of the samples. The probe was \textit{s}-polarized in order to exclude interference effects between the pump and probe pulses. The diameter of the pump beam exceeded that of the probe beam by a factor of 2.6. The intensity of the reflected probe beam was measured with a silicon photodiode and a lock-in amplifier.

\subsubsection*{Fit of experimental data:}

To quantitatively investigate the evolution of the relaxation dynamics, we fit the temperature dependence of the normalized reflectivity change according to the relation
\begin{align}
    \frac{\Delta R}{R} &=C_1 \cdot e^{-\gamma_1 t }+C_2 \cdot e^{-\gamma_2 t}+\Delta_0.
    \label{eq:M1}
\end{align}
\noindent
Depending on the temperature range in question, we may choose the relaxation rates $\gamma_{1,2}$ as real or complex, the amplitudes $C_{1,2}$ as positive, negative, or complex, and the offset $\Delta_0$ as nonzero or zero, as discussed in the following.

\begin{figure}
    \centering
    \includegraphics[width=\textwidth]{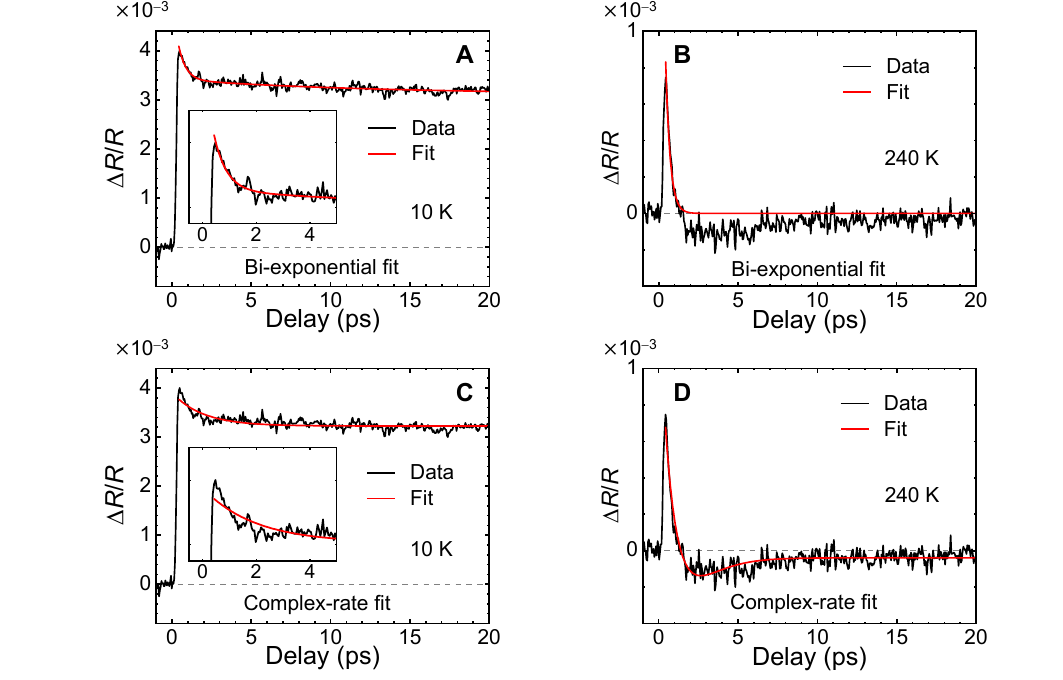}
    \caption{Normalized, pump-induced reflectivity $\Delta R/R$ data and fit results of Eq.~(\ref{eq:M1}) for the biexponential and the complex-rate scenarios discussed in the supplementary text. {\bf a}, {\bf b}, biexponential fit with real and positive values $C_{1,2}$, $\gamma_{1,2}$, and $\Delta_0$. It reproduces the low-temperature data well, but fails to reproduce the negative part of $\Delta R/R$ at high temperature. \textbf{c, d,} Data and fit results with complex values $C_{1,2}=C_0 e^{\pm i\varphi_0}$ and $\gamma_{1,2}=\gamma' \pm i \gamma''$. This model describes the data appropriately at high temperature, but fails at low temperature, indicating a qualitative change of the dynamics at an intermediate temperature $T^*$. The insets in \textbf{a} and \textbf{c} are zoomed-in views for better visualization of the fit quality.}
    \label{fig:biexponential+complex}
\end{figure}

At low temperature, the time dependence of $\Delta R/R$ in the main text suggests a biexponential relaxation, and we choose $C_{1,2}>0$ and $\gamma_{1,2}$ as real, so that the latter represent relaxation rates with $\tau_\text{slow}=1/\gamma_1$, $\tau_\text{fast}=1/\gamma_2$ as the associated relaxation times. The offset $\Delta_0>0$ describes the reflectivity change associated with the long-lasting insulator-metal transition resulting from the Stoner band splitting in the ferromagnetic phase; see Fig.~\ref{fig:OD_spectrum} and the discussion in the main text. Typical fit results are presented in Fig.~\ref{fig:biexponential+complex}, (A) and (B). They show that the biexponential model perfectly describes the relaxation behavior of $\Delta R/R$ at low temperature (panel \textbf{a}) but fails to reproduce the reflectivity behavior at high temperature (panel \textbf{\rm b}), where $\Delta R/R$ exhibits a negative regime.

\begin{figure}
    \centering
    \includegraphics[width=0.5\textwidth]{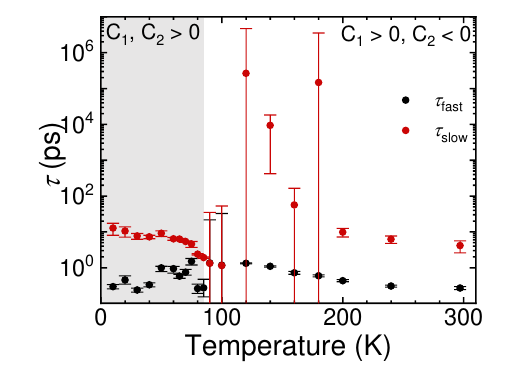}
    \caption{Verification of biexponential fit with $C_{1,2}>0$ below $T^*$ and $C_1>0$, $C_2<0$ above $T^*$ for pump fluence $P=100~\mu J/cm^2$. The fit is unable to consistently describe the negative contribution to the reflectivity as exponential behavior according to the unphysically large relaxation time and enormous error bars. This behavior is independent of the chosen laser pump fluence.}
    \label{fig:biexponential_negative}
\end{figure}

In the main text, we list a variety of processes one might associate with a negative contribution to the reflectivity. In the discussion in the Supplementary Text, we exclude all these from being responsible for the behavior seen in Fig.~\ref{fig:time_trace}D, but as an additional verification, we nevertheless attempted to fit such processes as a biexponential behavior with one positive and one negative contribution. We do this by choosing $C_1<0$ and $C_2>0$, keeping $C_{1,2}$ and $\gamma_{1,2}$ real as before. The results are displayed in Fig.~\ref{fig:biexponential_negative}. We see that the fit is unable to consistently describe the negative contribution to the reflectivity as exponential behavior, which is expressed by the unphysically large relaxation time and enormous error bars. The fit fails especially in the important temperature range where the negative contribution emerges, a behavior that is, furthermore, unaffected by the chosen pump fluence. Therefore, the negative-amplitude model ($C_1<0$) must be discarded not only on physical grounds as above, but also from the point of view of data fitting.

Instead, we emulate the region with $\Delta R/R<0$ by assuming complex values for $C_{1,2}$ and $\gamma_{1,2}$. Specifically, we set $C_{1,2}=C_0 e^{\pm i\varphi_0}$ and $\gamma_{1,2}=\gamma' \pm i \gamma''$, which ensures that the resulting observable $\Delta R/R<0$ in Eq.~(\ref{eq:M1}) represents a real value, as requested. We furthermore set $\Delta_0=0$ because there is no insulator-metal transition and Stoner splitting in the high-temperature paramagnetic regime; see Fig.~\ref{fig:OD_spectrum}. This leads to only two adjustable parameters, $\gamma'$ and $ \gamma''$, since the amplitude $C_0$ is absorbed in the normalization of the experimental signal and the phase $\varphi_0$ is determined by the position of the zero crossing of the experimentally observed time traces (see, e.g.~Fig.~\ref{fig:biexponential+complex}D). From Fig.~\ref{fig:biexponential+complex}, (C) and (D) we see that the complex-rate model fits the experimental data very well at high temperature but, as expected, fails at low temperature. 

\begin{figure}
    \centering
    \includegraphics[width=0.5\textwidth]{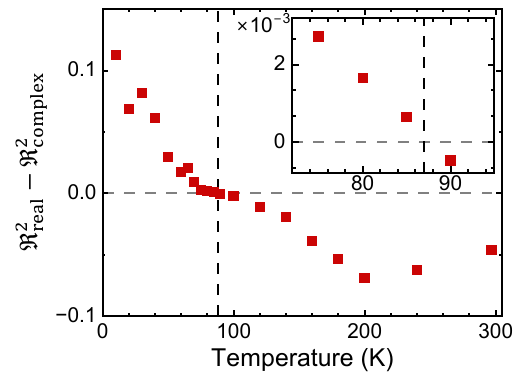}
    \caption{Difference in fit quality according to Eq.~(\ref{eq:M2}) between the biexponential and complex-rate scenarios. Better suitability of the former or latter are indicated by positive or negative values of $\mathfrak{R}^2_\text{real}-\mathfrak{R}^2_\text{complex}$, respectively. We see that with increasing temperature, a transition from biexponential to complex-rate behavior occurs with the zero crossing at $(87.8\pm 1.6)$~K in agreement with the fit based on Fig.~\ref{fig:exp_rates}B.}
    \label{fig:R2}
\end{figure}

To quantify the feasibility of the biexponential and complex-rate fits and evaluate the transition temperature $T^*$ between the respective regimes, we assess the quality of the fit by the parameter.
\begin{equation}
  \mathfrak{R}^2=1-\frac{\sum_i\left(y_i-f_i\right)^2}{\sum_i\left(y_i-\bar{y}\right)^2},
   \label{eq:M2}
\end{equation}
where $y_i$ represent the measured data at the observation times $t_i$, $f_i$ is the predicted value by the fitting model and $\bar{y}$ is the average of the measured data. Thus, the larger $\mathfrak{R}^2$, the better the model describes the experimental data. Figure~\ref{fig:R2} shows the difference $\mathfrak{R}^2_{\rm real}-\mathfrak{R}^2_{\rm complex}$ between the real- and complex-valued relaxation models. We see that within the possible range of $\mathfrak{R}^2_{\rm real}-\mathfrak{R}^2_{\rm complex}$ between $+1$ and $-1$ the biexponential model with positive amplitudes is more appropriate at low temperature, whereas the complex-rate model consistently describes the data better at high temperature. This behavior is characteristic for a NHPT, as discussed in detail in the main text. The transition temperature of the NHPT, where the change from real to complex relaxation occurs, is derived as $T^*=(84.4\pm 2.6)$~K from the extrapolation of the temperature dependence of $\gamma''$ in Fig.~\ref{fig:exp_rates}B. This value is in agreement with the zero crossing of $\mathfrak{R}^2_{\rm real}-\mathfrak{R}^2_{\rm complex}$ at $T^*=(87.8\pm 1.6)$~K in Fig.~\ref{fig:R2}. On a final note, the complex relaxation rate might be associated with an oscillation. However, because of its pronounced overdamping, we find this connotation misleading and, therefore, do not use it.

\subsubsection*{Theoretical modeling:} 

We use the Lindblad formalism \cite{Petruccione2010} to derive the dynamical equations for the numbers of bright (spin-0) and dark (spin-1) excitons as well as for the $z$-component of the Eu $4f$ spins from the microscopic Hamiltonian of EuO. Here, acoustic phonons, which assist in bright-dark exciton transformations, comprise a thermal bath at the cryostat temperature $T$ which induces effective temperature-dependent spin-flip couplings $\Gamma^{\pm}(T)$ between excitons and the Eu $4f$ magnetic moments, with $\Gamma^+(T)\ll\Gamma^-(T)$. In the paramagnetic phase, this leads, after linearization, to a set of two coupled dynamical equations for the deviations $\Delta n_{\rm b}(t)$, $\Delta n_{\rm d}(t)$ of the bright or dark exciton densities from their peak values after the pump pulse, respectively. 
\begin{equation}
    \frac{\rm d}{dt}
    \begin{pmatrix}
        \Delta n_{\rm b} \\
        \Delta n_{\rm d} 
    \end{pmatrix} = \chi (T)
    \begin{pmatrix}
        \Delta n_{\rm b} \\
        \Delta n_{\rm d} 
    \end{pmatrix}, \qquad
     \chi(T)= \begin{pmatrix}
     \gamma_{\rm b} & \omega_1 \\
     -\omega_2 & \gamma_{\rm d}
    \end{pmatrix}
    \label{eq:dynamical_eq1}\ .
\end{equation} 
The expressions for $\gamma_{\rm b}$, $\gamma_{\rm d}$, and $\omega_1$, $\omega_2$ are derived in the following section. The eigenvalues $\gamma_{1,2}$ of this set of equations are given in Eq.~(\ref{eq:eigenvalue}), with ${\omega_0}^2=\omega_1\omega_2$. They depend on temperature through $\Gamma^{\pm}(T)$ and change from real- to complex-valued at the critical temperature $T^*$ of the NHPT, implying the square-root dependence of the complex relaxation parameter $\gamma'' (T) = {\rm Im} [\gamma_{1,2}] \propto \sqrt{T-T^*}$. In the ferromagnetic phase, the temperature-dependent, average magnetization of the lattice of Eu $4f$ moments as well as the dynamics of their deviations from the average and the spin-dependent dynamics of the polarized excitons have to be taken into account, leading to a set of four dynamical equations; see Supplementary Materials. The appearance of the magnetization with a critical exponent $\alpha\approx 1/2$ \cite{Mairoser2010,Arnold2008} leads to the cusp feature in the relaxation times at the Curie temperature $T_{\rm C}$ seen in Fig.~\ref{fig:theo_rates}A which also appears in the experimental data, Fig.~\ref{fig:exp_rates}A.

\subsubsection*{Theory of exciton dynamics:}

The dynamical variables of the photoexcited EuO system are the density of bright (spin-0) and dark (spin-1) excitons, as well as the orientation $S_{i,z}$ of the Eu $4f$ magnetic moments $\vec S_i$ at the lattice site $i$, $ S_{i,z} =\pm 1/2,~\pm 3/2, \dots , \pm S,~ S=7/2$. The spin-singlet excitons are represented by bosonic creation and destruction operators, $a_{0}^{\dagger}$, $a_0^{\phantom{\dagger}}$, the triplet excitons by $a_{\pm 1}^{\dagger}$, $a_{\pm 1}^{\phantom{\dagger}}$ for spin quantum numbers $m=\pm 1$, respectively. The effective Hamiltonian for these excitations reads
\begin{align}
H_{\rm ex} = \sum_{m=\pm 1}\hslash\Omega_m^{\phantom{\dagger}} a_{m}^{\dagger}a_{m}^{\phantom{\dagger}} -
J \sum_{\langle i, j \rangle} \vec{S}_i \cdot \vec{S}_j + 
\sum_{k} \varepsilon_k b_k^{\dagger} b_k^{\phantom{\dagger}} +
J_{df} \sum_{m=\pm ,~k}(b_k + b_k^{\dagger}) 
\left[
 a_m^{\dagger}a_0^{\phantom{\dagger}} S^{-m} +
 a_0^{\dagger}a_m^{\phantom{\dagger}} S^{m}
\right]\ , \label{eq:hamiltonian}
\end{align},
where $\hslash\Omega_{m}>0$ is the excitation energy of a dark exciton ($m=\pm 1$) relative to a bright exciton ($m=0$). $J$ is the ferromagnetic-polaron exchange coupling between the Eu $4f$ moments on a Heisenberg lattice, and the last term describes the transfer between bright and dark excitons. It is induced by the spin-exchange coupling $J_{df}$ between the electron spin in a $5d(t_{2g})$ orbital as a constituent of an exciton and the $4f$ moments on neighboring Eu sites \cite{Mauger1977}. Here and in the following, $S^+$ and $S^-$ represent the corresponding 4f-spin raising and lowering operators, respectively, and we suppress the index of neighboring Eu lattice sites for simplicity of notation. Note that the triplet dark-exciton state with $m=0$ is not accessed by the spin-flip processes and therefore does not enter the dynamics. The bright-dark exciton transitions are assisted by absorption or emission of phonons with dispersion $\varepsilon_k$, described by phonon creation and destruction operators $b_k^{\dagger}$. The phonon system comprises a thermal Markovian bath at the cryostat temperature $T$ that we treat by the Lindblad formalism. This leads, after lengthy, but straightforward operator algebra \cite{Petruccione2010}, \cite{Oeztuerk2021} to the Lindblad master equation for the density matrix $\hat \rho(t)$ of the coupled exciton-Eu $4f$-moment system,
\begin{align}
\frac{d\hat\rho(t)}{dt} = i[\hat{\rho}(t), H_0] &+ p(t)~L[a_0^{\dagger}]~\hat{\rho}(t) +
R ~ L[a_0]~ \hat{\rho}(t) 
\label{eq:master} \\
&+ \sum_{m=\pm 1} \left\{
\Gamma^+ ~ L[a_m^{\dagger}a_0^{\phantom{\dagger}} S^{-m}]~\hat{\rho}(t) +
\Gamma^- ~ L[a_0^{\dagger}a_m^{\phantom{\dagger}} S^{m}]~\hat{\rho}(t)
\right\} \ . \nonumber
\end{align} 
Here, the Lindblad superoperator $L$ of an operator $A$ acting on the density matrix is defined as
\begin{align}
L[A]~ \hat{\rho} = 
A\hat{\rho}A^{\dagger} - (A^{\dagger}A\hat{\rho} + \hat{\rho}A^{\dagger}A )/2 \ .
\label{eq:superop}
\end{align}
The right-hand side of Eq.~(\ref{eq:master}) thus describe, in order of appearance, the coherent von-Neumann time evolution, the excitation of bright excitons by the external laser pulse $p(t)$, the radiative decay of bright excitons with amplitude $R$, and the phonon-assisted bright-dark and dark-bright exciton transformations mediated by raising and lowering the Eu $4f$ spin, described by the spin raising and lowering operators $S^+$ and $S^-$, respectively. For reasons of energy conservation (rotating wave approximation), the former process is accompanied by phonon absorption, the latter by phonon emission, since $\Omega_{\pm 1} > 0$. The bare system Hamiltonian $H_0$ consists of the first two terms of Eq.~(\ref{eq:hamiltonian}) and the bright-dark/dark-bright exciton transformation terms without coupling to phonons. Eq.~(\ref{eq:master}) contains the time-dependent, pump-pulse rate $p(t)$ whose temporal integral represents the pump fluence $P$ of the experiments. Integrating the thermal phonon bath from the Hamiltonian $H_{\rm ex}$ leads to effective temperature-dependent couplings for dark-exciton excitation and de-excitation processes in the master Eq.~(\ref{eq:master}), respectively, 
\begin{align}
\Gamma_m^+(T) &= \frac{{J_{df}}^2}{2} N_{\rm ph}(\Omega_{m})\,B(\Omega_{m}) 
\label{eq:Gamma+}\\
\Gamma_m^-(T) &= \frac{{J_{df}}^2}{2} N_{\rm ph}
(\Omega_{m})\,[1+B(\Omega_{m})]  \ , 
\label{eq:Gamma-}
\end{align}
where $B(\Omega_{m})= 1/[\exp{(\hslash\Omega_{m}/k_{\rm B}T)}-1]$ is the Bose-Einstein distribution function at temperature $T$ of the phonon bath, with $k_{\rm B}$ the Boltzmann and $\hslash$ the reduced Planck constant, and $N_{\rm ph}(\Omega_{m})$ the phonon spectral density per unit cell at the dark-exciton excitation energy $\hslash\Omega_{m}$, $m=\pm 1$. The excitation and de-excitation amplitudes thus obey the detailed-balance or Kennard-Stepanov relation \cite{Stepanov1957}, 
$\Gamma_m^+/\Gamma_m^- = \exp(-\hslash \Omega_{m}/k_{\rm B}T)$. Recall that the bright-dark exciton transitions are mediated by the magnetic coupling to the Eu 4f magnetic moments (not by the THz pulse), which is significantly larger than the thermal energy of phonons. Hence, $\hslash\Omega_{m}/k_{\rm B}T\gg 1$ at the relevant temperatures, and we have $\Gamma_m^+\ll\Gamma_m^-$. 

The statistical expectation value of a physical quantity $\hat{X}$ is defined as $\langle \hat{X} \rangle:={\rm tr}\left\{\hat{\rho}(t)\hat{X}\right\}$. We use the short-hand notation $n_m = \langle a^{\dagger}_{m}a^{\phantom{\dagger}}_{m} \rangle$ for the expectation values of the exciton density operators, $m=0,\,\pm 1$, and $S_z = \langle \hat{S}_z \rangle$ for the expectation value of the $z$ component of the $4f$ spin. Inserting Eqs.~(\ref{eq:master}) and (\ref{eq:superop}), we thus obtain the non-linear rate equations for the dynamics following the initial pump pulse,
\begin{align}
     \dfrac{dn_{0}}{dt}  = &- R n_{0} 
      - \Gamma_{1}^{+} \langle S^{+} S^{-} \rangle  \left( 1+  n_{1}   \right)   n_{0}   
      - \Gamma_{-1}^{+} \langle S^{-} S^{+} \rangle  \left( 1+  n_{-1}  \right)   n_{0}   \label{eq:n0}\\
      &\hspace*{0.9cm}+ \Gamma_{1}^{-} \langle S^{-} S^{+} \rangle  \left( 1+  n_{0}   \right)   n_{1}  
      + \Gamma_{-1}^{-} \langle S^{+} S^{-} \rangle  \left( 1+  n_{0}  \right)   n_{-1}   \nonumber
     \\
     \dfrac{dn_{1}}{dt}      = & 
     - \Gamma_{1}^{-} \langle S^{-} S^{+} \rangle  \left( 1 +    n_{0}  \right)    n_{1}   \label{eq:n1}
     + \Gamma_{1}^{+} \langle S^{+} S^{-} \rangle  \left( 1 + n_{1}  \right) n_{0}  
     \\
     \dfrac{dn_{-1}}{dt}       = &
     - \Gamma_{-1}^{-} \langle S^{+} S^{-} \rangle \left( 1 +   n_{0}    \right)    n_{-1}  
      + \Gamma_{-1}^{+} \langle S^{-} S^{+} \rangle \left( 1 +    n_{-1}    \right)    n_{0}    \label{eq:n-1}      
     \\
     \dfrac{dS_{z}}{dt}       = & 
     - \Gamma_{1}^{+} \langle S^{+} S^{-} \rangle \left( 1 +   n_{1}   \right)    n_{0}  
     + \Gamma_{-1}^{+} \langle S^{-} S^{+} \rangle \left( 1 +   n_{-1}   \right)   n_{0}    \label{eq:Sz} \\
     & - \Gamma_{-1}^{-} \langle S^{+} S^{-} \rangle \left( 1 +  n_{0}   \right)   n_{-1}    +  \Gamma_{1}^{-} \langle S^{-} S^{+} \rangle  \left( 1 +   n_{0}    \right)  n_{1}    
     \ ,\nonumber
\end{align}
where $\langle S^{\pm}S^{\mp} \rangle = S(S+1) - \langle {\hat{S}_z} ^{\ 2} \rangle \pm \langle \hat{S}_z\rangle $. Note that coherent time evolution due to $H_0$ does not contribute because bright-dark exciton conversion without phonon assistance is effectively forbidden by energy conservation ($\Omega_{\pm 1} > 0 $) and the commutator involving the number operators contained in $H_0$ vanishes under the trace. The dynamical equations (\ref{eq:n0})-(\ref{eq:Sz}) may be simplified by defining the total density of dark excitons as $n_{\rm d}=n_{1}+n_{-1}$ and their spin polarization as $m_{\rm d}=n_{1}-n_{-1}$ and observing that the total magnetization, consisting of the spin polarizations of the $4f$ moments and of the dark excitons, is conserved by these dynamics, that is, $dm_{\rm d}/dt = -dS_z/dt$.   

In order to analyze the relaxation behavior of the system, the rate equations (\ref{eq:n0})--(\ref{eq:Sz}) are expanded to linear order in the dynamical variables about their peak values, $\overline{n_m}$, $m=0,\pm 1$, after excitation by the initial pump-laser pulse. This leads to a set of four coupled linear rate equations for the deviations of the exciton densities and of the $4f$-spin orientation $S_z(t)$ from their respective initial values.   

\begin{table*}[b!]
\centering
\begin{tabular}{|c|c|c|c|c|c|c|c|c|c|}
\hline
$R$ & $\overline{n_{\rm b}}$ & $\overline{n_{\rm d}}$ & $\Gamma^{-}/R$ & $\Gamma^{+}/R$ \\ \hline
0.1 THz & 0.07 & $0.25\,n_{\rm b}$ & $0.006 + 6.9 \times 10^{-5} \cdot k_{\rm B}T/(\hbar R)$ & $\Gamma^{-} / 100$ \\ \hline
\end{tabular}
\\[0.3cm]
\caption{Parameter values used in the paramagnetic phase to generate Fig.~4 of the main article and Fig.~\ref{fig:time_trace_theory}. Values are chosen to reproduce the experimentally observed NHPT temperature of $T^*\approx 85$ K.}  
\label{tab:parameters}. 
\end{table*}

\begin{figure}
    \centering
   \includegraphics[width=\textwidth]{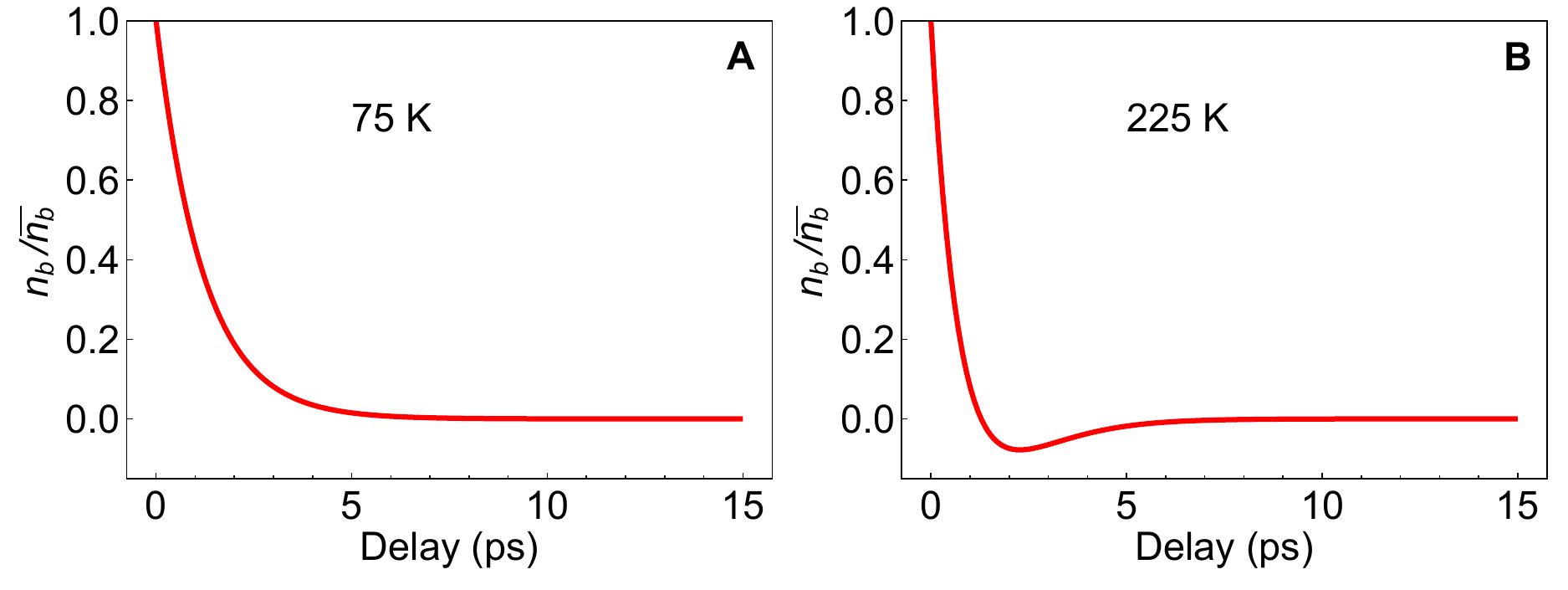}
    \caption{Theoretical time traces of the bright-exciton population $n_{\rm b}(t)$, computed from the dynamical equations (\ref{eq:dynamical_eq2}) {\bf (A)} on the biexponential side ($75$~K) and {\bf (B)} on the complex-rate side ($225$~K) of the NHPT, respectively. The data are normalized to the initial value $\overline{n_{\rm b}} = n_{\rm b}(0)$. The parameter values used for the calculation are given in Table~\ref{tab:parameters}.}
    \label{fig:time_trace_theory}
\end{figure}

In the paramagnetic phase, the $4f$ magnetization before the pump pulse is $S_z (0)=0$, and for symmetry reasons, the populations of up-spin and down-spin dark excitons are equal,  $n_{1} = n_{-1}$. It then follows from Eq.~(\ref{eq:Sz}) that $S_z(t) =0$ at all times. Hence, the transverse $4f$ spin correlator $C_{\perp}=\langle S^{+}S^{-}\rangle = \langle S^{-}S^{+}\rangle = S(S+1)-\langle {\hat{S}_z}^{\ 2}\rangle = {\rm const.}$ and $\Gamma_{1}^{\pm}=\Gamma_{-1}^{\pm}\equiv \Gamma^{\pm}$. 
The linearized dynamical equations then simplify to a set of two coupled equations for the bright and dark exciton densities, respectively. 
\begin{align}
    \frac{\rm d}{dt}
    \begin{pmatrix}
        \Delta n_{\rm b} \\
        \Delta n_{\rm d} 
    \end{pmatrix} &= \chi (T)
    \begin{pmatrix}
        \Delta n_{\rm b} \\
        \Delta n_{\rm d} 
    \end{pmatrix} \label{eq:dynamical_eq2} \\
     \chi(T)&= \begin{pmatrix}
        -R + 2C_{\perp} \left[\Gamma^{-} \overline{n_{\rm d}} - \Gamma^{+} (1 + \overline{n_{\rm d}})\right]\quad &
          \quad 2C_{\perp} \left[\Gamma^{-} (1+ \overline{n_{\rm b}}) - \Gamma^{+} \overline{n_{\rm b}}\right]\\
         -C_{\perp} \left[\Gamma^{-} \overline{n_{\rm d}} -\Gamma^{+}(1+\overline{n_{\rm d}})\right] & \quad
         C_{\perp} \left[ \Gamma^{+} \overline{n_{\rm b}} -\Gamma^{-} (1 + \overline{n_{\rm b}})\right]
    \end{pmatrix},
    \nonumber
\end{align} 
where we used the notation for the deviation of the bright and dark exciton densities from their respective initial values, $\Delta n_{\rm b}(t) = n_0(t)- \overline{n_0}$ and $\Delta n_{\rm d}(t) =n_{\rm d}(t)-\overline{n_{\rm d}}$, and $\overline{n_{\rm b}}$, $\overline{n_{\rm d}}$ are the corresponding initial values after the pump pulse. The dynamical matrix $\chi(T)$ has the eigenvalues 
\begin{equation}
  \gamma_{1,2}(T)= \left[\gamma_{\rm b} + \gamma_{\rm d} \pm \sqrt{(\gamma_{\rm b}-\gamma_{\rm d})^2 -4 \omega_0^2}~\right]/2, 
  \label{eq:eigenvalue2}
\end{equation}
with
\begin{align}
\gamma_{\rm b}&= -R + 2C_{\perp} \left[\Gamma^{-} \overline{n_{\rm d}} - \Gamma^{+} (1 + \overline{n_{\rm d}})\right]
\nonumber \\ 
\gamma_{\rm d}&=C_{\perp} \left[ \Gamma^{+} \overline{n_{\rm b}} -\Gamma^{-} (1 + \overline{n_{\rm b}})\right]
\label{eq:matrix_elements}\\ 
\omega_0&=C_{\perp} \sqrt{2 \left[\Gamma^{-} (1+ \overline{n_{\rm b}}) - \Gamma^{+} \overline{n_{\rm b}}\right]  \left[\Gamma^{-} \overline{n_{\rm d}} -\Gamma^{+}(1+\overline{n_{\rm d}})\right] }  \ . \nonumber
\end{align}
The eigenvalues depend on temperature through $\Gamma^{-}(T)$ and $\Gamma^{+}(T)$  [c.f.\ Eqs.~(\ref{eq:Gamma+}), (\ref{eq:Gamma-})] and change from real to complex at the critical temperature $T^*$ of the exceptional point (EP). Expanding the argument of the square root in Eq.~(\ref{eq:eigenvalue2}) to linear order of $(T-T^*)$, it follows that the imaginary relaxation parameter behaves as $\gamma''={\rm Im}(\gamma_{1,2})\sim \sqrt{T-T^*}$, as shown in Fig.~4b of the main text.
The parameter values used for the numerical evaluations are shown in Table~\ref{tab:parameters}. Fig.~\ref{fig:time_trace_theory} shows typical time traces of bright and dark exciton densities in the phases of biexponential and complex-rate relaxation of the NHPT, respectively. 

In the ferromagnetic phase, the $4f$ magnetization acquires a nonvanishing static value $M(T)$ in addition to its time-dependent variation $\langle S_z\rangle =M+\langle\Delta S_z\rangle$, and the up-spin and down-spin dark-exciton populations are no longer equal, $n_{1} \neq n_{-1}$. In this case, the full set of four dynamical equations (\ref{eq:n0})--(\ref{eq:Sz}), after linearization, must be solved simultaneously. According to ferromagnetic mean-field theory, which is well controlled for the large Eu $S=7/2$ spins, the equilibrium magnetisation has a square-root temperature dependence near the Curie temperature, $M(T)=M_0\sqrt{1-T/T_{\rm C}}$, which leads to the cusp signature in the relaxation times at $T_{\rm C}$, as shown in Fig.~4a of the main text.

\subsection*{Supplementary Text}

\subsubsection*{Discussion of alternative mechanisms:}

A variety of common physical processes are known to be capable of causing a negative dynamic response of the reflectivity of the type seen in Fig.~\ref{fig:time_trace}D. Among the multitude of effects that might come to mind are oscillations induced by coupling to coherent lattice vibrations \cite{Sun2017}, Auger recombination \cite{But2019, Sharma2024}, band filling effects in semiconductors \cite{Sabbah2002, Prabhu2004, Cicco2011, Cicco2020}, temperature or strain-dependent band shifts \cite{Jnawali2020}, three-temperature-model relaxation \cite{Shim2020}, lattice heating, excitation of optical phonons, or paramagnetic excitations \cite{Miyamoto2018}. We do not further discuss phenomena occurring on different time scales \cite{Sabbah2000}, in different frequency ranges [THz] \cite{Ye2018,Bonetti2022}, or in different material phases [like superconductivity] \cite{Mihailovic2014}.

Let us first take a closer look at some of these processes. We begin with oscillations induced by coupling to coherent lattice vibrations \cite{Sun2017}. This effect can be discarded because in the rock-salt structure of EuO, the relevant optical phonons are at a much higher frequency than the frequency corresponding to the slow $\lesssim 10$~ps dynamics observed in our experiments. Specifically, the optical phonon frequency is approximately 5.4~THz \cite{Goian2020}, whereas the observed value of $\gamma''$ corresponds to $0.02$~THz in our case; therefore, optical phonons can be excluded as the origin of the observed dynamics. Next, a negative dynamical response may generally also be caused by Auger recombination \cite{But2019, Sharma2024}. In a metal or doped semiconductor, a photoexcited electron could recombine with a hole in a radiationless way by exciting another electron to higher energy, thus reducing the number of carriers contributing to the reflectivity. However, in insulating EuO above 69~K, Auger recombination would be of second order in the carrier density and thus proportional to the square of the pump fluence, in contrast to our experimental and theoretical derivation of a square-root dependence. In addition, we can exclude band filling effects in semiconductors, which occur at fluences about 100 times higher than in our case \cite{Sabbah2002, Prabhu2004, Cicco2011, Cicco2020}. Photocarrier thermalization and lattice heating in EuO occur on a time scale of $\gg 10$~ps \cite{Masakazu2015}, which lies beyond the time window relevant to our observations.

We can work our way through a number of other mechanisms in the same way. This is not necessary, however, because beyond the point-by-point discussion of the different physical effects, there is an argument that calls into question the totality of such explanations for the negative dynamic response: None of the aforementioned effects is able to explain (i) the very specific temperature and pump-fluence dependencies of the relaxation rates and (ii) the presence of a critical temperature $T^*$ for the onset of these dependencies --- all clearly expressed by Fig.~\ref{fig:exp_rates}. In contrast, our model of a NHPT explains aspects (i) and (ii) in a seamless way, and in addition, several other aspects of our experimental observations.

\end{document}